\documentclass[a4paper,twocolumn,english,prl]{revtex4}
\usepackage[T1]{fontenc}
\usepackage[latin1]{inputenc}
\usepackage{amsmath}
\usepackage{graphicx}
\usepackage{amssymb}

\makeatletter
\date{\today}
\usepackage{ae,aecompl}

\usepackage{bm}

\usepackage{babel}
\makeatother
\begin{document}
\title{Many-body quantum dynamics of polarisation squeezing in optical fibre} 

\author{Joel F. Corney and Peter D. Drummond}

\affiliation{ARC Centre of Excellence for Quantum-Atom Optics, School of Physical
Sciences, The University of Queensland, Brisbane, QLD 4072, Australia}

\email{corney@physics.uq.edu.au}

\author{Joel Heersink, Vincent Josse, Gerd Leuchs and Ulrik. L. Andersen}

\affiliation{Institut f\"ur Optik, Information und Photonik, Max-Planck Forschungsgruppe,
Universit\"at Erlangen-N\"urnberg, Erlangen, Germany}

\begin{abstract}
We report new experiments that test quantum dynamical predictions of polarization squeezing for ultrashort photonic pulses in a birefringent fibre, including all relevant dissipative effects. This exponentially complex many-body  problem is solved by means of a stochastic phase-space method. The squeezing is calculated and compared to experimental data, resulting in excellent quantitative agreement. From the simulations, we identify the physical limits to quantum noise reduction in optical fibres. The research represents a significant experimental test of first-principles time-domain quantum dynamics in a one-dimensional interacting Bose gas coupled to dissipative reservoirs. 
\end{abstract}
\maketitle

The nonlinear optical response of standard communications fibre
provides a straightforward and robust method 
\cite{Drummond_etal93} for squeezing the quantum noise always present
in laser light to below the vacuum noise level. This feature allows us to design quantum dynamical experiments
\cite{SizmannLeuchs,expt} operating in a very nonclassical regime where 
highly entangled states can be readily produced,
even in many-body regimes involving $10^8$ interacting particles. 
 The squeezing is sensitive to many-body 
photon-photon interactions, as well as to additional dissipative and 
thermal effects \cite{DrummondCorney01}. Such complications have affected all 
prior experiments and have so far prevented  quantitative 
agreement between theory and experiment. 

Here we report on quantitative comparisons of first-principles 
simulations with experimental measurements on the propagation of quantum states in optical fiber.  The excellent agreement, over a wide range of initial conditions, is unprecedented for direct predictions from ab-initio treatments of many-body quantum time-evolution.  The approach we use has potential applications in many other areas of science,
especially to dynamical experiments with ultra-cold atoms and nanotechnology.

Photons in a nonlinear fibre are an implementation of the
famous one-dimensional attractive Bose gas model\cite{Mcguire} in quantum field theory.
Fibre squeezing experiments thus provide a substantial opportunity to carry out an experimental test of the predictions of many-body quantum mechanics for dynamical time-evolution. Such a test requires the
same  ingredients as did Galileo's famous tests of classical dynamics using
an inclined plane\cite{Galileo}: one needs a known initial condition, a well-defined cause of dynamical evolution, and
accurate measurements. All of these essential features are  present in our experiments. The initial condition is a coherent\cite{Glauber} photonic state provided by a well-stabilized pulsed laser.  The Kerr nonlinearity in silica fibre corresponds to a localized
(delta-function) interaction between the photons\cite{Yurke}.  Quantum-limited phase-sensitive measurements have been developed in optics that are able to detect quantum fluctuations at well below the vacuum noise level\cite{Walls83LoudonKnight87,DrummondFicek04}.  

Even though the simplest model of a 1D interacting Bose gas has exactly soluble energy eigenvalues, the many-body initial value
problem  still remains intractably complex. Expanding the coherent initial state directly in terms of more than $10^{100}$ relevant eigenstates (more than the number of particles in the universe) is simply not practical. For such reasons, Feynman suggested\cite{Feynman} that the calculation of dynamical
quantum time-evolution would be impossible without
special `quantum computing' hardware, which is not available at present. Even worse,
there are additional complications occurring due to coupling to phonons in the silica fibre medium. 

To deal with the exponential quantum complexity in this dissipative system, using existing computers, we employ quantum phase-space methods based on the work of Wigner\cite{Wigner} and Glauber\cite{Glauber}.  With modifications to treat non-classical fields\cite{+P}, phase-space methods are able to calculate quantum correlations in a dynamical system without recourse to hydrodynamic approximations or linear response theory.  They were successfully
used for the original prediction of quantum squeezing in optical fibres,\cite{Carter_etal87DrummondCarter87}, which was qualitatively
verified in several laboratories\cite{RosenbluhShelby91,Drummond_etal93}.
Recently, photon-number squeezing was simulated \cite{Werner98}
and compared with experiment \cite{Schmitt_etal98}. However, the
cause of squeezing degradation was unclear\cite{Fiorentino_etal01}. By using an improved
experimental configuration\cite{expt}, and including all relevant dissipative effects in the simulations, we have been able to quantitatively test and verify the phase-space predictions. 

Physically, the Kerr effect that generates the squeezing can be viewed as producing an intensity-dependent refractive index, which distorts an initially symmetric phase-space
distribution into an ellipse
Because of energy conservation, however, the variance in the amplitude remains constant. Thus
the squeezing cannot be detected directly in amplitude or intensity
measurements.

\begin{figure}
\includegraphics[%
  width=1.0\columnwidth]{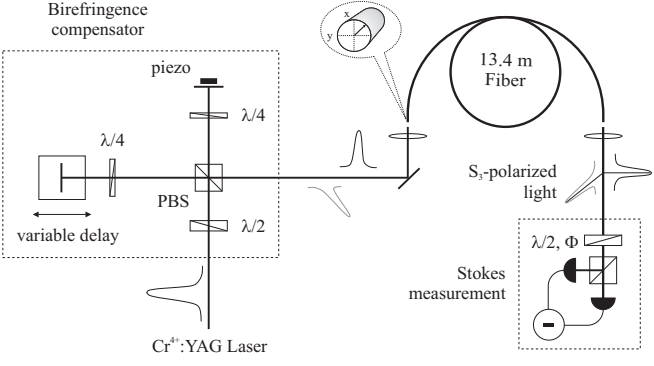}
  \caption{Experimental detail for the efficient generation of polarization squeezed states.  Pairs of coherent, 74 fs FWHM
sech-shaped laser pulses at 1.5 $\mu$m are coupled into a birefringent fiber.  A  pre-fiber variable delay ensures that the pulses overlap upon exitting the fiber. A relative phase shift of $\frac{\pi}{2}$
is maintained by means of a piezo-electric crystal in a feedback loop, resulting in circularly or $S_3$-polarized 
light. The squeezing measurement is performed 
by use of a half-wave plate, polarizing beam splitter and two balanced 
detectors. Although the squeezing is broad-band, it is 
sampled at a frequency of $17.5~{\textrm{MHz}}$ to avoid technical noise.}
\end{figure}

Although laser homodyne measurements\cite{DrummondFicek04} can
be used for these phase-sensitive measurements, they require a local oscillator,
which is impractical for the high-intensity laser pulses used here.
Instead, we employ an interferometric method known as polarization
squeezing\cite{expt}, wherein two multi-photon femptosecond pulses that are initially identical in
amplitude propagate along the mutually orthogonal polarization axes
$x$ and $y$ of a polarization-maintaining fiber. The two output
pulses are then combined on a half-wave plate at an angle $\Phi$.  For the appropriate phase-space rotation angles $\theta=4\Phi$, the squeezing
or antisqueezing can be detected in the number difference.
Because both pulses undergo similar evolution, most excess thermal
noise is common to both modes and is thus cancelled. An exception
here is any noise induced by birefringent, or depolarizing, effects,
which will be unique to each mode.  The experimental set-up is illustrated in Fig.~1.

Because the experiment involves ultrashort pulses, we use photon-density operators for each polarisation mode of the optical field: $\widehat{\Psi}_{x}(t,z)$ and $\widehat{\Psi}_{y}(t,z)$
that include a range of spectral components $\widehat{a}(t,k)$:
\begin{eqnarray}
\widehat{\Psi}_{\sigma}(t,z) & \equiv & \frac{1}{\sqrt{2\pi}}\int dk\,\widehat{a}_{\sigma}(t,k)e^{i(k-k_{0})z+i\omega_{0}t}\,,\end{eqnarray}
where $\sigma=x,y$ labels the polarisation, $t$ is time, $z$ is propagation distance and $k$ is wave-number.  The wavenumber and frequency of the carrier wave are given by $k_0$ and $\omega_0$.  The commutation relations of the field operators
are $\left[\widehat{\Psi}_{\sigma}(t,z),\widehat{\Psi}_{\sigma'}^{\dagger}(t,z')\right]=\delta(z-z')\delta_{\sigma\sigma'}$. 

To describe the polarisation squeezing, we define generalized Stokes operators:\begin{eqnarray}
\widehat{S}_{0}\equiv\widehat{N}_{xx}(L)+\widehat{N}_{yy}(L), && \widehat{S}_{1}\equiv\widehat{N}_{xx}(L)-\widehat{N}_{yy}(L),\nonumber \\
\widehat{S}_{2}\equiv\widehat{N}_{xy}(L)+\widehat{N}_{yx}(L), && \widehat{S}_{3}\equiv i\widehat{N}_{yx}(L)-i\widehat{N}_{xy}(L),
\end{eqnarray}
where $L$ is the fibre length and $\widehat{N}_{\sigma\sigma'}(z)=\int dt\widehat{\Psi}_{\sigma}^{\dagger}(t,z)\widehat{\Psi}_{\sigma'}(t,z)$.  $\widehat{S}_{0}$ corresponds to the combined intensity, $\widehat{S}_{1}$
gives the number difference between the $x$ and $y$ modes, and $\widehat{S}_{2}$
and $\widehat{S}_{3}$ are operators sensitive to the relative phase difference.  In the Stokes 
measurement after the fiber, we measure a given Stokes parameter orthogonal to the bright excitation in $S_3$.
This, the dark $\widehat{S}_1$-$\widehat{S}_2$ plane, is scanned by rotating a half-wave plate to measure: $\widehat{S}_{\theta}=\cos(\theta)\widehat{S}_{1}+\sin(\theta)\widehat{S}_{2}.$
Following from the Stokes operator commutation relations\cite{KitagawaUeda93Korolkova_etal02}, 
the number difference is squeezed if the variance of $\widehat{S}_{\theta}$
is below, for $S_3$-polarized light,\begin{eqnarray}
\Delta^{2}\widehat{S}_{\theta}  <  \left| \left\langle \widehat{S}_{3}\right\rangle \right|  =  \left| \left\langle \widehat{S}_{0}\right\rangle \right|.\end{eqnarray}

To simulate the evolution of the photon density $\widehat{\Psi}_{\sigma}(t,z)$,
we employ a quantum model of a radiation field propagating along a fused
silica fibre, including the $\chi^{(3)}$ nonlinear responses of the material
and the nonresonant coupling to phonons\cite{DrummondCorney01,CarterDrummond91}. The
phonons act as a reservoir with finite-time correlations (non-Markovian) that generates additional,
delayed nonlinearity, as well as spontaneous and thermal noise. Classical, low-frequency phase noise can be neglected, becase it is common to both polarization modes, unless there is a depolarising component. The phonon spectrum
is based on the experimentally determined Raman gain for pure fused
silica $\alpha^{R}(\omega)$\cite{StolenLeeJain84}. Because of the fiber birefringence, 
the two polarization components do not temporally overlap for most of 
the fiber length, so we neglect the cross-polarization component of the Raman gain.

To simplify the theoretical description, we use a propagative frame with rescaled time $\tau\equiv(t-z/v)$, rescaled propagation distance
$\zeta\equiv z/z_{0}$ and dimensionless photon-flux field $\widehat{\phi}_{\sigma}\equiv\widehat{\Psi}_{\sigma}\sqrt{vt_{0}/\overline{n}}$. Here $t_{0}$ is the pulse duration, $z_{0}\equiv t_{0}^{2}/|k''|$
is the dispersion length and  $2\overline{n}$
is the photon number in a fundamental ${\rm sech}(\tau)$ soliton pulse.  The fibre-dependent parameters are pulse velocity $v$, and dispersion parameter $k''$.

From the multimode Hamiltonian for the full system\cite{DrummondCorney01}, we obtain Heisenberg equations of motion for the quantum fields. The equations for the phonon field are integrated to derive quantum Langevin
equations for the photon-flux field
\begin{eqnarray}
\frac{\partial}{\partial\zeta}\widehat{\phi}_{\sigma}(\tau,\zeta)  =  \frac{i}{2}\frac{\partial^{2}}{\partial\tau^{2}}\widehat{\phi}_{\sigma}(\tau,\zeta)+i\widehat{\Gamma}_{\sigma}(\tau,\zeta)\widehat{\phi}_{\sigma}(\tau,\zeta)
&&\nonumber \\ 
+ i\int_{-\infty}^{\infty}d\tau'h(\tau-\tau')\widehat{\phi}_{\sigma}^{\dagger}(\tau',\zeta)\widehat{\phi}_{\sigma}(\tau',\zeta)\widehat{\phi}_{\sigma}(\tau,\zeta)&&,
\label{eq:RamanOperator}\end{eqnarray}
where the nonlinear response function $h(\tau)$ includes 
both the instantaneous electronic response and the Raman response
determined by the gain function $\alpha^{R}(\omega)$\cite{StolenLeeJain84,DrummondCorney01,CarterDrummond91}.
The correlations of the reservoir fields are\begin{eqnarray}
\left\langle \widehat{\Gamma}_{\sigma}^{\dagger}(\omega',\zeta')\widehat{\Gamma}_{\sigma'}(\omega,\zeta)\right\rangle  & = & \frac{\alpha^{R}(|\omega|)}{\overline{n}}\left[n_{\mathrm{th}}(|\omega|)+\Theta(-\omega)\right]\nonumber \\ &  &
 \times\delta(\zeta-\zeta')\delta(\omega-\omega')\delta_{\sigma\sigma'},
 \label{RamanOperatorCorrelations}\end{eqnarray}
where $n_{{\rm th}}$ is the temperature-dependent Bose distribution
of phonon occupations. The Stokes ($\omega<0$) and anti-Stokes ($\omega>0$)
contributions to the Raman noise are included by means of the
step function $\Theta$. These equations can easily be modified to
include coupling to gain and absorption reservoirs, but this is 
unnecessary for the short distances in these experiments.

The exact quantum dynamics can been simulated using the positive-P ($+P$)
phase-space representation\cite{Carter_etal87DrummondCarter87}. However, for
large photon number $\overline{n}$ and short propagation distance
$L$, it is known that the $+P$ method gives squeezing predictions
in agreement with a truncated Wigner phase-space method\cite{Wignersim},
which allows faster calculations. We have chosen the latter method
to reduce computational overheads.  This method maps a field operator to a stochastic field: $\widehat{\phi}_{\sigma}(\zeta,\tau)\rightarrow\phi_{\sigma}(\zeta,\tau).$
Stochastic averages involving this field then correspond to symmetrically
ordered correlations of the quantum system. Because of the symmetric-ordering
correspondence, quantum effects enter via vacuum noise. The Kerr effect
merely amplifies or diminishes this noise in a phase-sensitive manner,
which makes the Wigner approach ideally suited for squeezing calculations.

Using the Wigner mapping, we obtain a Raman-modified stochastic nonlinear
Schr\"{o}dinger equation for the photon flux that is of exactly
the same form as Eq.~(\ref{eq:RamanOperator})\cite{DrummondCorney01,CarterDrummond91}.
The correlations of the Raman noise fields $\Gamma_{\sigma}$ are also of the same form as their operator equivalent Eq. (\ref{RamanOperatorCorrelations}), except that the step function is replaced by a constant ${1}/{2}$. Because of the symmetrically ordered mapping, the Stokes and anti-Stokes
contributions to the Wigner Raman noise are identical.  
The initial condition of the stochastic field is the mean coherent level plus fluctuations that correspond to vacuum noise: 
\begin{eqnarray}
\left\langle \Delta\phi_{\sigma}(\tau,0)\,\Delta\phi_{\sigma'}^{*}(\tau',0)\right\rangle  & = & \frac{1}{2\overline{n}}\delta(\tau-\tau')\delta_{\sigma\sigma'}\,.\end{eqnarray}

\begin{figure}
\includegraphics[%
  width=0.50\columnwidth]{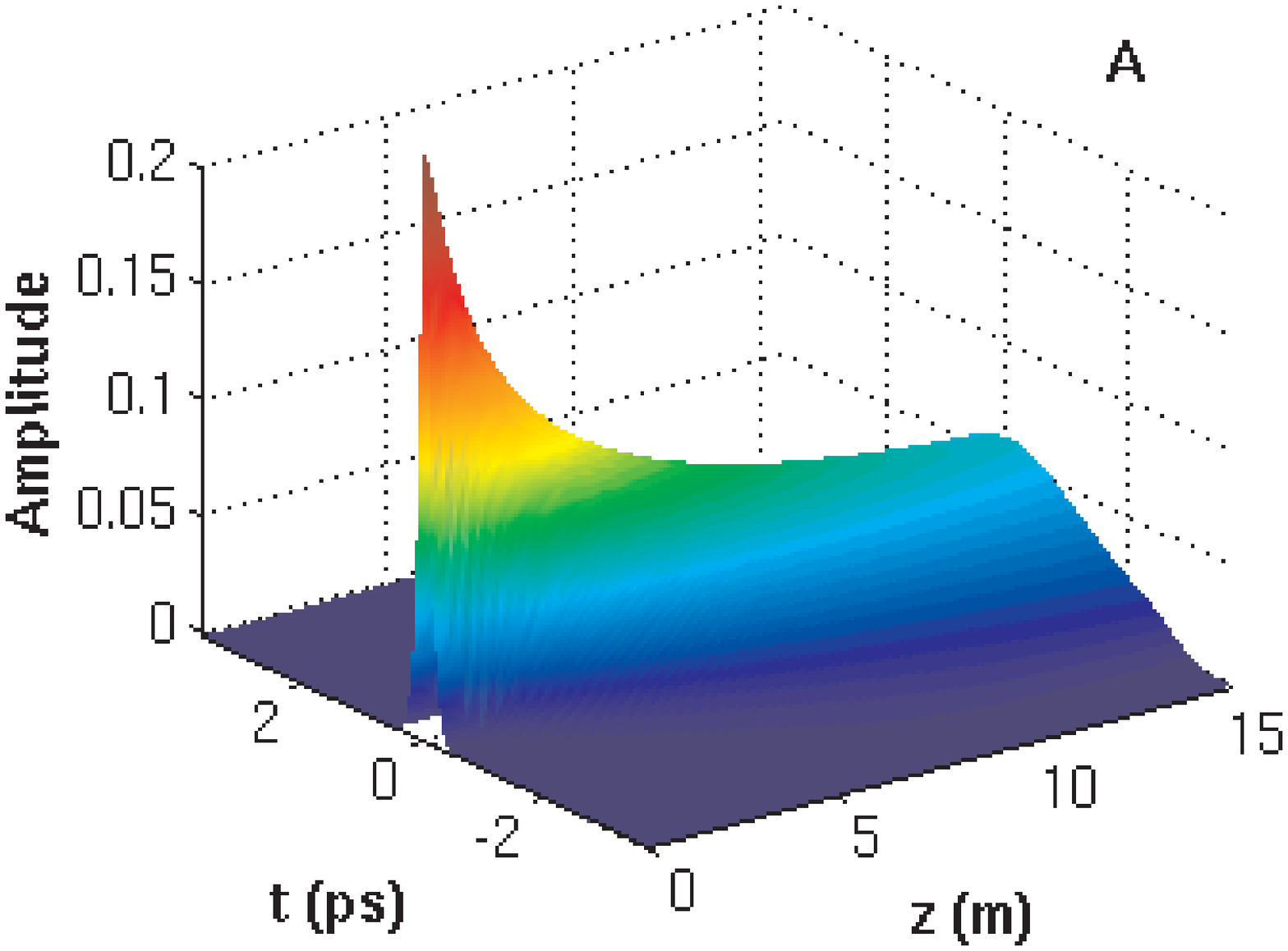}~\includegraphics[%
  width=0.50\columnwidth]{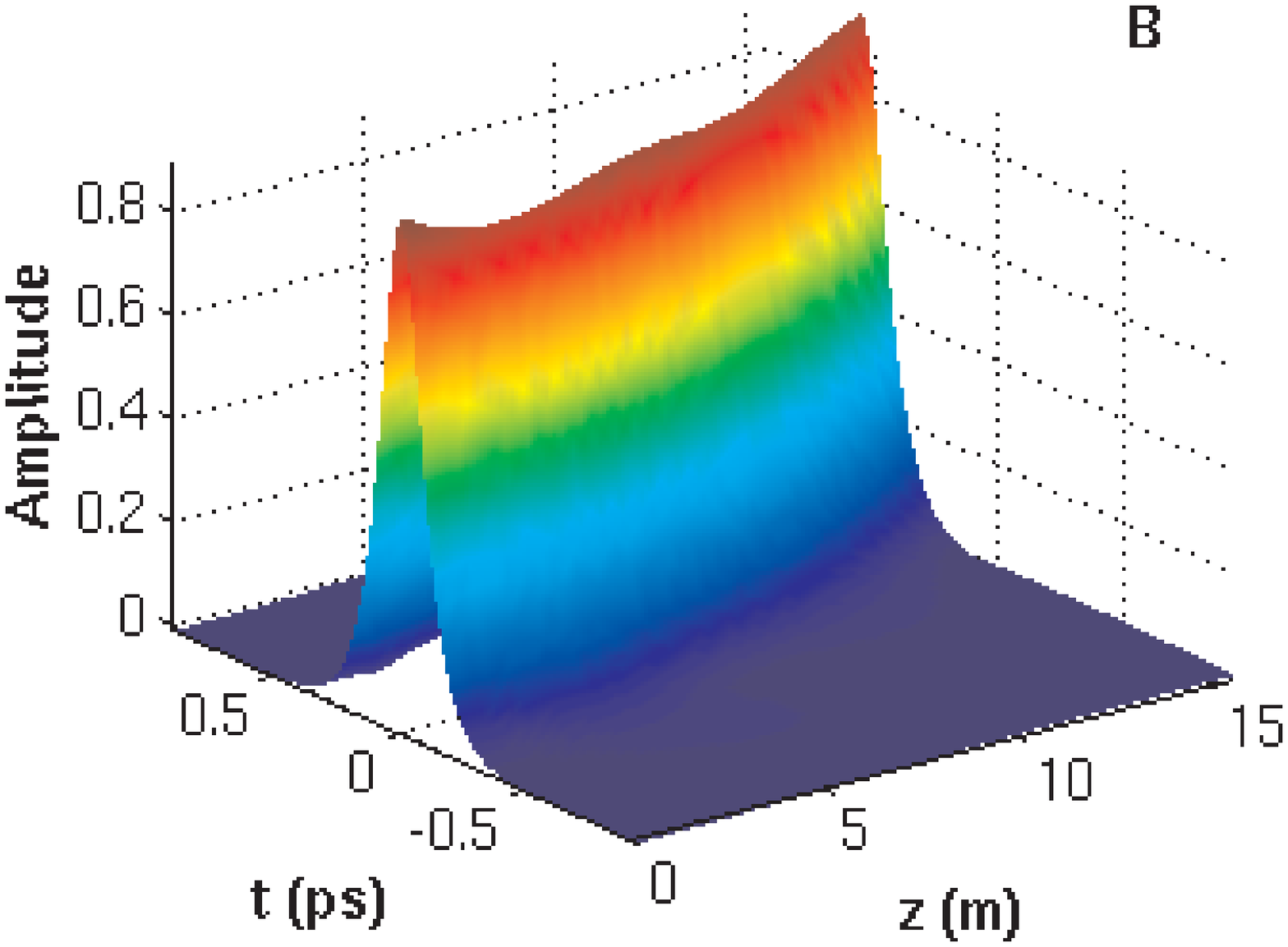}
  \caption{Propagation of (A) $E=4.8$~pJ and (B)
$E=53.5$~pJ pulses, with initial width $t_{0}=74$~fs.}
\end{figure}

Simulations of the pulse propagation reveal the importance of including
the system's multimode nature, which affect small-amplitude and intense
pulses in different ways. As Fig.~2 shows,
the evolution of the amplitude profile of a weak pulse is dominated
by dispersion. In contrast, an intense pulse reshapes into a soliton,
whose subsequent evolution reveals the effect of the Raman self-frequency
shift. The range of input pulse energies in the experiment includes
both these extremes.

\begin{figure}
\includegraphics[%
  width=0.50\columnwidth]{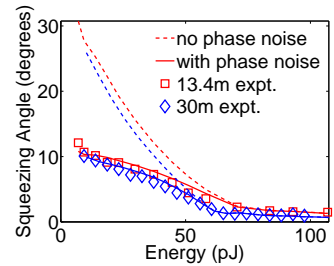}
  \caption{Phase-space rotation angle
$\theta$ versus total input energy $E$. Squares and diamonds give
the experimental results for the $L_{1}=13.4$~m and $L_{2}=30$~m
fibres, respectively. Continuous (dashed) lines give the simulation
results with (without) excess phase noise. Other parameters are $t_{0}=74$~fs,
$z_{0}=0.52$~m, $\overline{n}=2\times10^{8}$, and $\lambda_{0}=1.51$~$\mu$m.}
\end{figure}

Fig.~3 gives the phase-space
rotation angle $\theta$ at which squeezing is observed at different input energies.
The simulations and experiments agree very well for the intense pulses,
but there is a divergence for weak pulses. Now classical phase noise has a relatively strong effect on the weaker pulses, because the antisqueezing produced by the Kerr effect
is then smaller and at a greater angle to the phase quadrature.  We therefore include the effect of excess phase noise by a nonlinear least-squares procedure.  A new relative noise variance is calculated from 
\begin{eqnarray}
\Delta^2\widehat{S}_\theta/\left< \widehat{S}_3\right> &=& \rho_p\sin^2(\theta_N)+\rho_s\cos^2(\theta_N-\theta_K) \nonumber \\ &&
+ \rho_a\sin^2(\theta_N-\theta_K),
\end{eqnarray}
where $\theta_K(E)$ are the angles for Kerr squeezing with input energy $E$, and where $\rho_s(E)$, $\rho_a(E)$ and  $\rho_p(E)$ are the relative variances of the Kerr squeezing, Kerr antisqueezing and phase noise.  The Kerr variances and angles $\theta_K(E)$ are taken from the simulation data, and the phase noise is assumed to depend linearly on pulse energy\cite{ShelDruCar90}.  The coefficient of the phase noise is determined by a nonlinear least-squares fit of the angles $\theta_N(E)$ that minimise the new variance to the measured squeezing angles.   The result gives an excellent fit to the experimental data, even though there is only one parameter for each fibre length. This parameter was larger for the longer fiber length, indicating that phase noise is largely a fiber-induced effect, such as depolarizing guided acoustic wave Brillouin scattering (GAWBS).  However, there also seems to be a smaller birefringent noise source outside the fiber.

\begin{figure}
\includegraphics[%
  width=0.50\columnwidth,
  keepaspectratio]{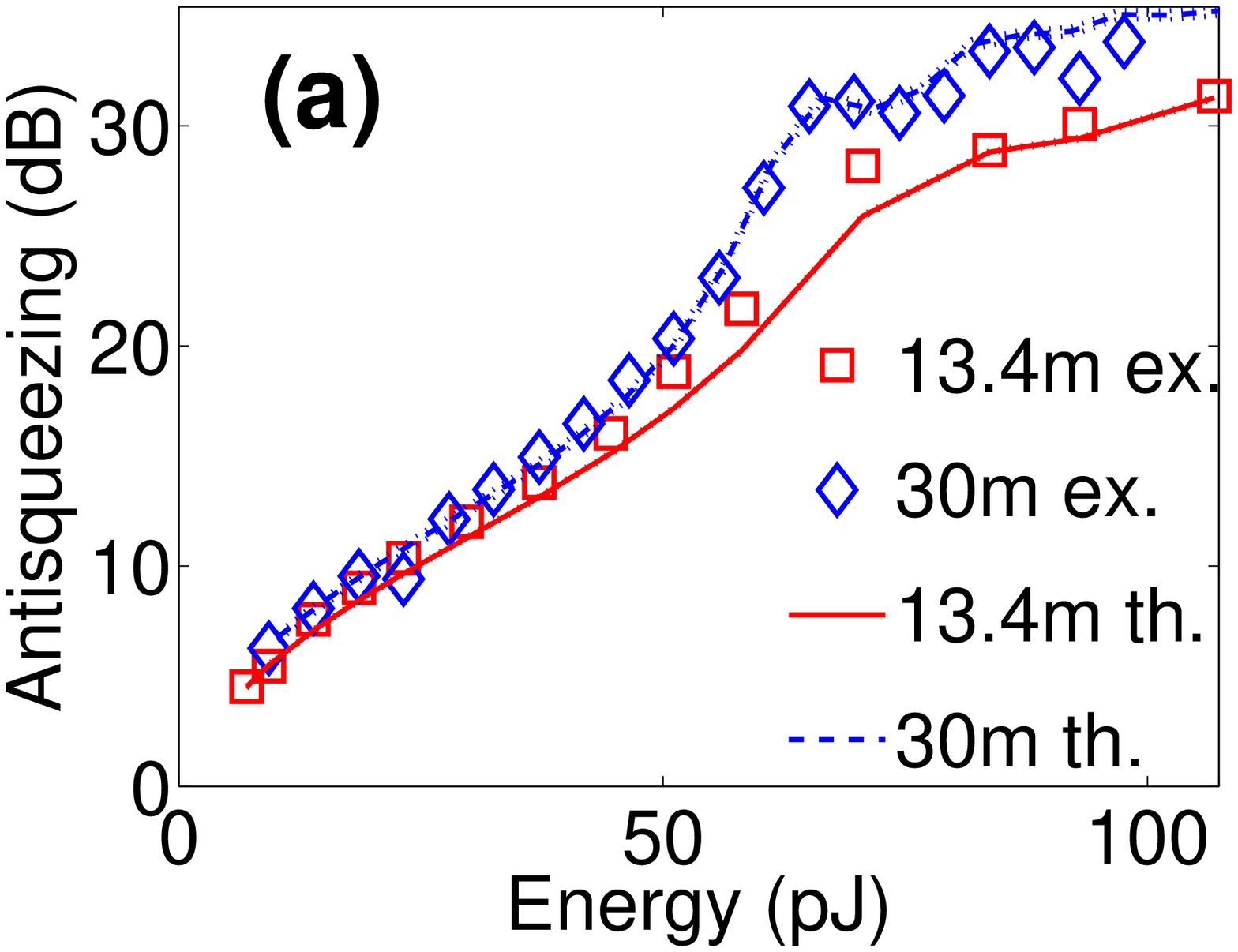}~\includegraphics[%
  width=0.50\columnwidth,
  keepaspectratio]{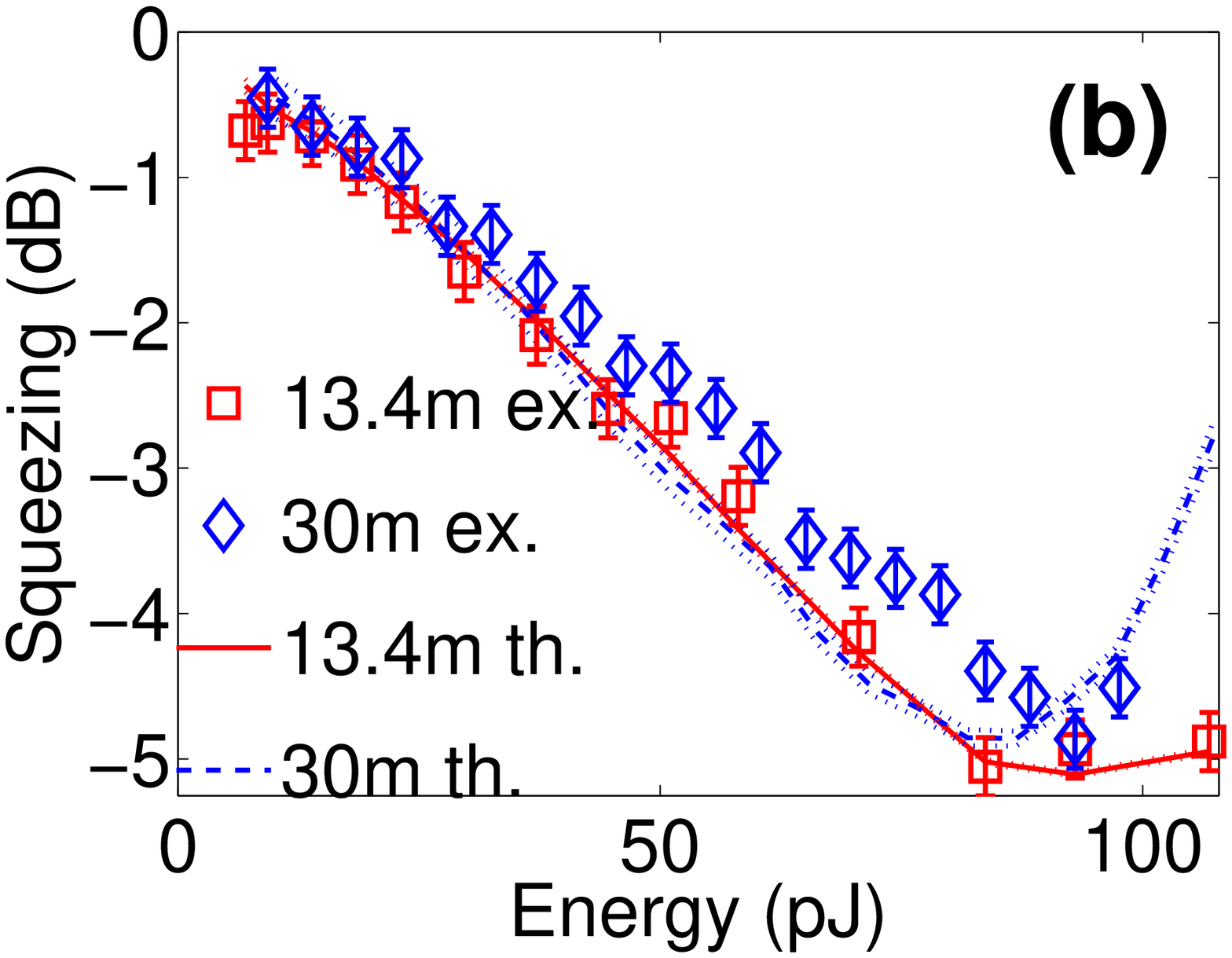}
  \caption{Raw antisqueezing (A) and squeezing (B)
for $L_{1}=13.4$~m (squares) and $L_{2}=30$~m
(diamonds) fibers. Error bars on the squeezing data indicate the uncertainty in the noise measurement; for the antisqueezing the error bars were to small to be plotted distinctly for the antisqueezing.  Solid and dashed lines show the simulation results with additional phase noise for $L_{1}$ and $L_{2}$, respectively. Dotted lines indicate the sampling error in the simulation results.}
\end{figure}

Antisqueezing and squeezing results are shown Fig.~4,
for 13.4~m and 30~m of fiber, with the excess phase noise included. 
The theoretical results for both squeezing and antisqueezing
closely match the experimental data, after estimated linear losses
of $24\%$ are taken into account. 
That the theoretical antisqueezing results match the experiment
indicates that the fit obtained from the rotation angle accounts for
all the excess classical noise. A deterioration of the squeezing is seen at higher intensity due to Raman effects, especially
for longer fiber lengths.

As an indication of the accuracy of the simulations, we predicted from
comparisons of the rotation angle and the anti-squeezing curves 
in theory and in experiment that
the 30~m fiber should have a $5\%$ larger core diameter than the shorter fiber. 
This was verified by fiber measurements.  There is a residual discrepancy in the highly sensitive
squeezing  measurements for the
30~m case at large pulse energies. This could be due to effects such as cross-polarization Raman scattering,
higher-order dispersion, or initial pulse-shape distortion.

In conclusion, the efficiency of the squeezing experiments
described here means that a comprehensive theoretical model must be
formulated and solved to quantitatively account for all observations.
Solving the model requires a first-principles simulation of quantum
time-evolution in a many-body system coupled to a non-Markovian reservoir.
We have achieved this by means of a Wigner phase-space representation,
thus obtaining excellent agreement between simulation and experiment
over a wide range of energies and fibre lengths. The simulations reveal
that Raman effects limit squeezing at high intensity and for longer
fibers and that depolarizing phase noise (i.e. GAWBS) limits squeezing
at low intensity.

\emph{Acknowledgements:}  We thank Ch.~Marquardt and H.~A.~Bachor 
for discussions.  For financial support, JFC and PDD thank the Australian Research Council, and JH, VJ, GL and ULA thank the EU project 
COVAQIAL and the Deutsche Forshungsgemeinschaft.

\end{document}